\documentclass[11pt,a4paper]{article}
\usepackage{amsmath,amssymb}
\usepackage{epsfig,graphicx}
\usepackage{cite}

\begin{document}

\title{
\begin{flushright}
{\normalsize Yaroslavl State University\\
             Preprint YARU-HE-07/01\\
             hep-ph/0701228} \\[10mm]
\end{flushright}
PLASMA INDUCED FERMION SPIN-FLIP\\ CONVERSION $f_L \to f_R + \gamma$} 

\author{A.~V.~Kuznetsov$^a$\footnote{{\bf e-mail}: avkuzn@uniyar.ac.ru},
N.~V.~Mikheev$^{a}$\footnote{{\bf e-mail}: mikheev@uniyar.ac.ru}
\\
$^a$ \small{\em Yaroslavl State (P.G.~Demidov) University} \\
\small{\em Sovietskaya 14, 150000 Yaroslavl, Russian Federation}
}

\date{}

\maketitle

\begin{abstract}
The fermion spin-flip conversion $f_L \to f_R + \gamma$ is considered, 
caused by the difference of the additional energies of the electroweak origin, 
acquired by left- and right-handed fermions (neutrino, electron) in medium. 
An accurate taking account of the fermion and photon dispersion in medium 
is performed. It is shown that the threshold arises in the process, caused 
by the photon (plasmon) effective mass. This threshold leaves no room for 
the so-called ``spin light of neutrino'' and ``spin light of electron'' 
in the real astrophysical situations. 
\\


\end{abstract}
 
\vfill

\begin{center}
 {\it Submitted to International Journal of Modern Physics A}
\end{center}


\newpage

\section{Introduction}
\label{sec:Introduction}

The most important event in neutrino physics of the last decades was 
the solving of the Solar neutrino problem, made in the unique experiment 
on the heavy-water detector at the Sudbury Neutrino 
Observatory.\cite{SNO:2001} -- \cite{SNO:2002b} 
This experiment, together with the atmospheric and the reactor neutrino 
experiments,\cite{SK:1998a} -- \cite{KamLAND:2003} 
has confirmed the key idea by 
B. Pontecorvo on neutrino oscillations.\cite{Pontecorvo:1957,Pontecorvo:1958} 
The existence of non-zero neutrino mass and lepton mixing is thereby 
established. 
The Sun appeared in this case as a natural laboratory for investigations 
of neutrino properties. 

There exists a number of natural laboratories, the supernova explosions, where 
gigantic neutrino fluxes define in fact the process energetics. 
It means that microscopic neutrino characteristics, such as the neutrino 
magnetic moment, the neutrino dispersion in an active medium, etc., would have 
a critical impact on macroscopic properties of these astrophysical events.

This is the reason for a growing interest to neutrino physics in an external 
active medium. In an astrophysical environment, the main medium influence 
on neutrino properties is defined by the additional 
energy $W$ acquired by a left-handed neutrino.\cite{Wolfenstein:1978} 
This additional energy gives, in part, an effective mass squared $m_L^2$ 
to the left-handed neutrino, 
\begin{equation}
m_L^2 = {\cal P}^2 = (E + W)^2 - {\bf p}^2 \,, 
\label{eq:mL}
\end{equation}
where ${\cal P}^\alpha$ is the neutrino four-momentum in medium, while 
$p^\alpha = (E,\, {\bf p})$ would form the neutrino four-momentum in vacuum, 
$E = \sqrt{{\bf p}^2 + m_\nu^2}$. 

Given a $\nu \nu \gamma$ interaction, 
the additional energy of left-handed neutrinos in medium opens new 
kinematical possibilities for the radiative neutrino transition: 
\begin{equation}
\nu \to \nu + \gamma \,.
\label{eq:nunugamma}
\end{equation}

It should be self-evident, that the influence of the substance on the photon 
dispersion must be taken into account, 
$\omega = |{\bf k}|/n$, where $n \ne 1$ is the refractive index. 

First, a possibility exists that the medium provides the condition $n > 1$ 
(the effective photon mass squared is negative, $m_\gamma^2 \equiv 
q^2 < 0$) which corresponds to the well-known 
effect\cite{Grimus:1993} -- \cite{Ioannisian:1997} of 
``neutrino Cherenkov radiation''.  
In this situation, the neutrino dispersion change under the medium influence 
is being usually neglected, because the neutrino dispersion is defined by the 
weak interaction while the photon dispersion is defined by the electromagnetic 
interaction. 

One more possibility could be formally considered when the photon 
dispersion was absent, and the process of the radiative neutrino transition
$\nu \to \nu + \gamma$ would be caused by the neutrino dispersion only. 
As the left-handed neutrino dispersion is only changed, transitions become 
possible caused by the $\nu \nu \gamma$ interaction with the neutrino 
chirality change, e.g. due to the neutrino magnetic dipole moment. 
Just this situation called the ``spin light of neutrino'' ($SL \nu$), 
was first proposed and investigated in detail in an extended series of 
papers.\cite{LobStuPLB03} -- \cite{StuJPA06} 

However, it is evident that in the analysis of this effect such an 
important phenomenon as plasma influence on the photon dispersion cannot be 
ignored. 
As will be shown below, this phenomenon closes the $SL \nu$ effect for all 
real astrophysical situations.\cite{Kuznetsov:2006a} -- \cite{Kuznetsov:2006d} 

In the recent papers\cite{0611100} -- \cite{0611128} the authors have extended 
their approach to the so-called ``spin light of electron'' ($SL e$), 
$e_L \to e_R + \gamma$. It will be shown that just the same mistake of ignoring 
the photon dispersion in plasma was repeated in these papers. 

It is interesting to note that it was not the first case when the plasma 
influence was taken into account for one participant of the physical process 
while it was not taken for other participant. The history is repeated. 

As E. Braaten wrote in Ref.~\cite{Braaten:1991}:

``In Ref.~\cite{Beaudet:1967}, it was argued that their calculation for the 
emissivities from photon and plasmon decay would break down at 
temperatures large enough that $m_\gamma > 2 \, m_e$, since the decay 
$\gamma \to e^+ e^-$ is then kinematically allowed. This statement, which 
has been repeated in subsequent 
papers,\cite{Dicus:1972} -- \cite{Itoh:1989} is simply untrue. 
The plasma effects which generate the photon mass $m_\gamma$ also generate 
corrections to the electron mass such that the decay $\gamma \to e^+ e^-$ 
is always kinematically forbidden.'' 

Thus, the 
authors\cite{LobStuPLB03} -- \cite{StuJPA06}, \cite{0611100} -- \cite{0611128} 
made the same mistake when they considered the plasma-induced additional 
neutrino or electron energy and ignored the effective photon mass $m_\gamma$ 
arising by the same reason.  

Consequently, the spin-flip processes $\nu_L \to \nu_R + \gamma$ and 
$e_L \to e_R + \gamma$ should be reanalysed with taking 
into account the photon dispersion in medium. Having in mind possible 
astrophysical applications, it is worthwhile to consider 
the astrophysical plasma as a medium, which transforms the photon into 
the plasmon, see e.g. Ref.~\cite{Braaten:1993} 
and the papers cited therein. 
Here, we perform a detailed analysis of the fermion spin-flip conversion 
$f_L \to f_R + \gamma$ with both the fermion dispersion and the photon 
dispersion in medium. 

\section{Kinematical Analysis for ``Spin Light of Neutrino''}
\label{sec:Analysis_SL_nu}

To perform the kinematical analysis, it is necessary to evaluate the scales 
of the values of the left-handed neutrino additional energy $W$ and of the 
photon (plasmon) effective mass squared $m_\gamma^2$.

The expression for this additional energy of 
a left-handed neutrino with the flavor $i = e, \mu, \tau$ was obtained in the 
local limit of the weak interaction,\cite{Notzold:1988} -- \cite{Nieves:1989} 
see also Ref.~\cite{Elmfors:1996}, and can be presented in the following 
form
%
\begin{eqnarray}
&&W_i = \sqrt{2} \, G_{\rm F} \left[
\left(\delta_{ie} - \frac{1}{2} + 2 \, \sin^2 \theta_{\rm W}\right) 
\left(N_e - \bar N_e \right) + \left(\frac{1}{2} 
- 2 \, \sin^2 \theta_{\rm W}\right) \left(N_p - \bar N_p \right) \right.
\nonumber\\
&& 
- \left. \frac{1}{2} \left(N_n - \bar N_n \right) 
+ \sum\limits_{\ell = e, \mu, \tau} \left(1 + \delta_{i \ell} \right)
\left(N_{\nu_\ell} - \bar N_{\nu_\ell} \right) 
\right] , 
\label{eq:EnuLgen}
\end{eqnarray}
where the functions 
$N_e, N_p, N_n, N_{\nu_\ell}$ are the number densities of background electrons, 
protons, neutrons, and neutrinos, and $\bar N_e, \bar N_p, \bar N_n, 
\bar N_{\nu_\ell}$ are the densities of the corresponding antiparticles.
To find the additional energy for antineutrinos, one should change the 
total sign in the right-hand side of Eq.~(\ref{eq:EnuLgen}). 

For a typical astrophysical medium, except for the early Universe, 
using the notations $N_p \simeq N_e = Y_e\, N_B, \, N_n \simeq  (1-Y_e)\, N_B, 
\, N_{\nu_e} = Y_{\nu_e}\, N_B$, where $N_B$ 
is the barion density, one obtains from Eq.~(\ref{eq:EnuLgen}):
\begin{eqnarray}
W_e &=& \frac{G_{\rm F} \, N_B}{\sqrt{2}} 
\left(3\, Y_e + 4\, Y_{\nu_e} - 1 \right) ,
\label{eq:EnuLe}\\
W_{\mu,\tau} &=& - \frac{G_{\rm F} \, N_B}{\sqrt{2}} 
\left(1 - Y_e  - 2\, Y_{\nu_e} \right) .
\label{eq:EnuLmu}
\end{eqnarray}
Here, $Y_{\nu_e}$ is not zero in the supernova core only, and describes the 
fraction of trapped electron neutrinos. 

In any medium except for the supernova core, the additional energy acquired 
by muon and tau left-handed neutrinos is always negative, because $Y_e < 1$. 
At the same time, the additional energy of electron 
left-handed neutrinos in these conditions becomes positive at $Y_e > 1/3$. 
And vice versa, the additional energy for electron antineutrinos is positive 
at $Y_e < 1/3$, while it is always positive for the muon and tauon 
antineutrinos. 
On the other hand, right-handed neutrinos and their antiparticles, left-handed 
antineutrinos, being sterile with respect to weak interactions, 
do not acquire an additional energy.

If $Y_{\nu_e}$ can be neglected, one readily obtains from 
Eq.~(\ref{eq:EnuLe}): 
\begin{equation}
W \simeq 6 \; {\rm eV}
\left(\frac{N_B}{10^{38} \, {\rm cm}^{-3}}\right) \left(3\, Y_e - 1 \right) ,
\label{eq:W}
\end{equation}
where the scale of the barion number density is taken, which is typical 
e.g. for the interior of a neutron star. 

On the other hand, a plasmon acquires in medium 
an effective mass $m_\gamma$ which is approximately constant at high energies. 
For the transversal plasmon, the value $m_\gamma^2$ is always positive, and 
is defined by the so-called plasmon frequency. 
In the non-relativistic classical plasma (i.e. for the solar interior) one has:
\begin{equation}
m_\gamma \equiv \omega_{\rm pl} = \sqrt{\frac{4 \pi \, \alpha \,N_e}{m_e}} \simeq 
4 \times 10^{2} \,{\rm eV}
\left(\frac{N_e}{10^{26} {\rm cm}^{-3}}\right)^{1/2}.
\label{eq:omega_pl_nr}
\end{equation}
For the ultra-relativistic dense matter one has:
\begin{equation}
m_\gamma^2 =  \frac{2 \, \alpha}{\pi} \left(\mu_e^2 + \frac{\pi^2}{3} \, T^2 \right),
\label{eq:omega_pl_r}
\end{equation}
where $\mu_e$ is the chemical potential of plasma electrons. 
For the case of cold degenerate plasma one obtains from 
Eq.~(\ref{eq:omega_pl_r}):
\begin{equation}
m_\gamma =  \sqrt{\frac{3}{2}} \; \omega_{\rm pl} 
= \left(\frac{2 \, \alpha}{\pi} \right)^{1/2} 
\left(3\, \pi^2 \, N_e \right)^{1/3} \simeq 
10^{7} \,{\rm eV}
\left(\frac{N_e}{10^{37} \, {\rm cm}^{-3}}\right)^{1/3}.
\label{eq:omega_pl_r_cold}
\end{equation}
In the case of hot plasma, when its temperature is the largest 
physical parameter, the plasmon mass is:
\begin{equation}
m_\gamma =  \sqrt{\frac{2 \, \pi \, \alpha}{3}} \; T
\simeq 
1.2 \times 10^{7} \,{\rm eV}
\left(\frac{T}{100 \, {\rm MeV}}\right).
\label{eq:omega_pl_h}
\end{equation}

One more physical parameter, a great attention was paid to in the $SL \nu$ 
analysis,\cite{LobStuPLB03} -- \cite{StuJPA06} 
was the neutrino vacuum mass $m_\nu$. 
As the scale of neutrino vacuum mass could not exceed essentially 
a few electron-volts, which 
is much less than typical plasmon mass scales for real astrophysical situations, 
see Eqs.~(\ref{eq:omega_pl_nr})-(\ref{eq:omega_pl_h}), it is reasonable 
to neglect $m_\nu$ in our analysis.

The energy and momentum conservation law for the radiative spin-flip neutrino 
transition $\nu_L \to \nu_R + \gamma$, with the additional left-handed 
neutrino energy included, has the form:
\begin{eqnarray}
E + W &=& E' + \omega \,,
\nonumber\\ 
{\bf p} &=& {\bf p'} + {\bf k} \,, 
\label{eq:mom_cons}
\end{eqnarray}
Thus, in accordance with~(\ref{eq:mL}), a simple condition for the 
kinematic opening of the process is:
\begin{equation}
m_L^2 \simeq 2 \, E \, W > m_\gamma^2 \,. 
\label{eq:mL>}
\end{equation}
This means that the process becomes kinematically opened when 
the neutrino energy exceeds the threshold value, 
\begin{equation}
E > E_0 = \frac{m_\gamma^2}{2 \, W} \,. 
\label{eq:E_0}
\end{equation}

This threshold behavior of the process 
$\nu_L \to \nu_R + \gamma$ in plasma becomes very clear if compared with 
the well-known process in vacuum. 
As is seen from Refs.~\cite{Grigoriev:2006a,Grigoriev:2006b}, the authors 
believe that they have shown in Refs.~\cite{GriStuTerGC05,GriStuTerPLB05} 
that ``for the case of high-energy neutrinos 
the matter influence on the photon dispersion can be neglected'' because 
``plasma is transparent for electromagnetic radiation on frequencies greater 
than the plasmon frequency''. It is rather naive consideration. 
Really, one can see that from the side of kinematics the discussed process 
$\nu_L \to \nu_R + \gamma$ in plasma is identical to the process 
$\bar\nu_e + e^- \to \tau^- + \bar\nu_\tau$, where the high-energy electron 
anti-neutrino scattered off the electron in rest, creates 
the $\tau$ lepton. 
The 4-momentum conservation law can be written in the form
\begin{equation}
p^\alpha + m_e \, u^\alpha = p '^\alpha + q^\alpha \,, 
\label{eq:tau_4-mom}
\end{equation}
where $p = (E, {\bf p} \,), \, p ' = (E ', {\bf p '})$ are the initial and final 
neutrino 4-momenta, $u$ is the 
4-vector of the electron velocity, which in its rest 
frame is $u^\alpha = (1,\, {\bf 0})$, $q = (\omega, {\bf k} \,)$ is the $\tau$ 
lepton 4-momentum. 
In the lab system, which is the electron rest frame, the energy and momentum 
conservation takes the form:
\begin{eqnarray}
E + m_e &=& E' + \omega \,,
\nonumber\\ 
{\bf p} &=& {\bf p'} + {\bf k} \,, 
\label{eq:tau_4-mom2}
\end{eqnarray}
which is just the same as Eq.~(\ref{eq:mom_cons}), up to the notations. 
Neglecting the neutrino masses, let us write down the Mandelstam $S$ variable 
in the lab system:
\begin{equation}
S = (p + m_e \, u)^2 = 2\, m_e \, E + m_e^2 \,. 
\label{eq:S1}
\end{equation}
On the other hand, in the center-of-mass frame one has:
\begin{equation}
S = (\omega + E')^2 = \left( \sqrt{m_\tau^2 + p'^2} + p' \right)^2 \geqslant m_\tau^2\,. 
\label{eq:S2}
\end{equation}
The threshold value for the initial neutrino energy arises from~(\ref{eq:S1}) 
and~(\ref{eq:S2}):
\begin{equation}
E \geqslant E_0 = \frac{m_\tau^2 - m_e^2}{2 \, m_e} \simeq \frac{m_\tau^2}{2 \, m_e} \,, 
\label{eq:E_0_tau}
\end{equation}
to be compared with Eq.~(\ref{eq:E_0}). The similarity is deliberately 
not accidental. Both inequalities are caused by the minimal value of the 
Mandelstam $S$ variable 
which is equal to the mass squared of the heavy particle in the 
final state, $m_\tau^2$ (or $m_\gamma^2$ in our case). At the same time, 
the mass of the initial electron in rest in the process 
$\bar\nu_e + e^- \to \tau^- + \bar\nu_\tau$ is kinematically identical to 
the additional left-handed neutrino energy $W$ in the process 
$\nu_L \to \nu_R + \gamma$ in plasma. 
Taking the approach of the authors,~\cite{LobStuPLB03} -- \cite{StuJPA06} 
one should forget about the threshold~(\ref{eq:E_0_tau}) and conclude that 
the process $\bar\nu_e + e^- \to \tau^- + \bar\nu_\tau$ is always open if only 
the medium (which is vacuum in this case) is transparent for $\tau$ leptons. 
 
Let us evaluate the threshold neutrino energies~(\ref{eq:E_0}) for different 
astrophysical situations. 

In the classical plasma, the threshold neutrino energy does not depend on 
density, and do depend on the chemical composition only:
\begin{equation}
E_0 \simeq \frac{Y_e}{3 Y_e - 1} \, \frac{4 \sin^2 \theta_{\rm W}\, 
m_{\rm W}^2}{m_e}\,. 
\label{eq:E_0_class}
\end{equation}
For the solar interior $Y_e \simeq 0.6$, and the threshold neutrino energy is
\begin{equation}
E_0 \simeq 10^{10} \,{\rm MeV} \,, 
\label{eq:p_0_sun}
\end{equation}
to be compared with the upper bound $\sim$ 20 MeV for the solar neutrino energies. 

For the interior of a neutron star, where $Y_e \ll 1$, 
the additional energy for neutrinos~(\ref{eq:EnuLe}), (\ref{eq:EnuLmu}) 
is negative, and the process $\nu_L \to \nu_R + \gamma$ is closed. 
On the other hand, there exists a possibility for opening the antineutrino 
decay. Taking for the estimation $Y_e \simeq 0.1$, one obtains from~(\ref{eq:W}) 
and~(\ref{eq:omega_pl_r}) the threshold value
\begin{equation}
E_0 \simeq 10^{7} \,{\rm MeV} \,, 
\label{eq:p_0_NS}
\end{equation}
to be compared with the typical energy $\sim$ MeV of neutrinos emitted 
via the URCA processes. 

For the conditions of a supernova core, $Y_e \sim 0.35, \, Y_{\nu_e} \sim 0.05$, 
the additional energy of left-handed electron neutrinos obtained 
from Eq.~(\ref{eq:EnuLe}) leads to:
\begin{equation}
E_0 \simeq 10^{7} \,{\rm MeV} \,, 
\label{eq:p_0_SN}
\end{equation}
to be compared with the averaged energy $\sim 10^{2}$ MeV of trapped neutrinos. 

Thus, the analysis performed shows that the radiative spin-flip neutrino 
transition $\nu_L \to \nu_R + \gamma$ is possible at ultra-high neutrino 
energies only. However, it should be obvious in this case, that the local 
limit of the weak interaction does not describe comprehensively the 
additional neutrino energy in plasma, and the non-local weak contribution 
must be taken into account. The analysis of this contribution was first 
performed for the conditions of the early 
Universe.\cite{Notzold:1988,Elmfors:1996}
In this case, the local weak contribution~(\ref{eq:EnuLgen}) is suppressed, 
because plasma is almost charge symmetric. 

In a general case, the non-local weak contribution into the additional 
neutrino energy in plasma, which is identical for both neutrinos and 
antineutrinos, can be presented in the form
\begin{equation}
\Delta^{(\rm{nloc})} W_i = 
- \frac{16 \, G_{\rm F} \, E}{3 \, \sqrt{2}} 
\left[ \frac{<E_{\nu_i}>}{m_Z^2} 
\left(N_{\nu_i} + \bar N_{\nu_i} \right)  
+ \delta_{ie} \, \frac{<E_e>}{m_W^2} \left(N_e + \bar N_e \right)
\right] , 
\label{eq:W_nloc}
\end{equation}
where $E$ is the energy of a neutrino propagating through plasma, $<E_{\nu_i}>$ 
and $<E_e>$ are the averaged energies of plasma neutrinos and electrons 
correspondingly. In a particular case of a charge symmetric hot plasma, 
this expressions reproduces the result of 
Refs.~\cite{Notzold:1988,Elmfors:1996}:
\begin{equation}
\Delta^{(\rm{nloc})} W_i = 
- \frac{7 \, \sqrt{2} \, \pi^2 \, G_{\rm F} \, T^4}{45} 
\left( \frac{1}{m_Z^2} + \frac{2 \, \delta_{ie}}{m_W^2} \right) E \, .
\label{eq:W_early}
\end{equation}
The minus sign in~(\ref{eq:W_early}) 
unambiguously shows that in the early Universe, in contrast 
to the neutron star interior, the process of the radiative spin-flip 
transition is forbidden both for neutrinos and antineutrinos. 

An analysis of the sum of the local and non-local weak 
contributions~(\ref{eq:EnuLgen}) and~(\ref{eq:W_nloc}) shows that adding 
of the latter leads in general to the decreasing of the the additional 
neutrino energy $W$ in plasma, i.e. to the increasing of the threshold 
energy~(\ref{eq:E_0}). Strictly speaking, one has to perform an analysis 
of the kinematical inequality~(\ref{eq:mL>}), which leads to the solving 
of the quadratic equation. As a result, there arises the window in the 
neutrino energies for the process to be kinematically opened, 
$E_0 < E < E_{\rm{max}}$, where $E_0$ and $E_{\rm{max}}$ are the lower and the 
upper limits connected with the roots of the above-mentioned quadratic 
equation, if they exist. For example, in the solar interior there is no window 
for the process with electron neutrinos at all, i.e. the transition  
$\nu_{eL} \to \nu_{eR} + \gamma$ is forbidden kinematically. 

Thus, the above analysis shows that the nice effect of the 
``spin light of neutrino'', unfortunately, has no place in real 
astrophysical situations because of the photon dispersion. 
The sole possibility for the discussed process 
$\nu_L \to \nu_R + \gamma$ to have any significance could be connected 
only with the situation when an ultra-high energy neutrino threads a star. 

\section{Kinematical Analysis for ``Spin Light of Electron''}
\label{sec:Analysis_SL_e} 

Similarly to a neutrino, an electron acquires in medium the additional energy 
depending on its helicity, due to the parity non-conserving weak 
interaction. To find this energy, one should consider the electron 
self-energy operator in medium. For a real electron, with taking account of 
the renormalization of the chiral mass and the wave function, 
this self-energy operator can be presented in the form
\begin{equation}
\Sigma_r  = \frac{1}{2} \, \hat{u} \left(V - A \, \gamma_5 \right)\,, 
\label{eq:Sigma}
\end{equation}
where $u^\alpha$ is the four-vector of plasma velocity which in its rest 
frame is $u^\alpha = (1,\, {\bf 0})$. The value $V$ is caused mainly by the 
electromagnetic interaction of an electron:
\begin{equation}
V \simeq V^{\rm{em}} = \frac{\alpha}{\pi \, E} \, \left( \mu_e^2 + \pi^2 T^2 \right).
\label{eq:Vem}
\end{equation}
Using~(\ref{eq:omega_pl_r}), the value $V$ can be expressed via the plasmon 
mass:  
\begin{equation}
V \simeq C \, \frac{m_\gamma^2}{E} \,, \quad 
C = \frac{3}{2} \, \frac{\mu_e^2 + \pi^2 T^2}{3 \mu_e^2 + \pi^2 \, T^2 } \,, 
\quad
\frac{1}{2} \leqslant C \leqslant \frac{3}{2}\,,
\label{eq:V_m_gamma}
\end{equation}
where the lower and the upper limits for $C$ correspond to the cases of cold 
and hot plasma. 

We do not present here the contribution into $V$ caused by the weak 
interaction, because in real astrophysical conditions the electromagnetic 
contribution $V^{\rm{em}}$ always dominates, except for the unphysical 
case of a pure neutron medium considered in 
Refs.~\cite{0611100} -- \cite{0611128}. 
It should be noted that even in the conditions of a cold neutron star, 
the fraction of electrons and protons cannot be exacly zero, 
$Y_e \gtrsim 0.01$.\cite{ShapiroTeukolsky} 
And even at such minimal value of $Y_e$ the electromagnetic contribution into 
$V$ dominates. 

As about the value $A$, it is caused by the weak interaction only and has 
the form: 
\begin{eqnarray}
&& A = \frac{G_{\rm F}}{\sqrt{2}} \bigg[
2 \left(1 - 4 \, \sin^2 \theta_{\rm W}\right) 
\left(N_e - \bar N_e \right) - \left(1 - 4 \, \sin^2 \theta_{\rm W}\right) 
\left(N_p - \bar N_p \right) 
\nonumber\\
&& + \left(N_n - \bar N_n \right) 
+ 2 \, \sum\limits_{\ell = e, \mu, \tau} \left(2 \, \delta_{\ell e} - 1 \right)
\left(N_{\nu_\ell} - \bar N_{\nu_\ell} \right) 
\bigg] 
\nonumber\\
&& - 
\frac{8 \, G_{\rm F} \, E}{3 \sqrt{2}} \left[ \frac{< \! E_e \! >}{m_Z^2} 
\left(N_e + \bar N_e \right) \left(1 - 4 \, \sin^2 \theta_{\rm W}\right) 
+ 4 \, \frac{< \! E_{\nu_e} \! >}{m_W^2} \left(N_{\nu_e} + \bar N_{\nu_e} \right)
\right] . 
\label{eq:A_gen}
\end{eqnarray}
Here, both the local and the non-local weak contributions are included. $E$ 
is the energy of an electron propagating through plasma, $<\! E_e\! >$ 
and $<\! E_{\nu_i}\! >$ are the averaged energies of plasma electrons and neutrinos 
correspondingly. 
For typical astrophysical medium, $\bar N_e \simeq \bar N_p \simeq \bar N_n 
\simeq \bar N_{\nu_\ell} \simeq 0$,
$N_p \simeq N_e = Y_e\, N_B, \, N_{\nu_e} \simeq Y_{\nu_e} N_B$, 
the expression~(\ref{eq:A_gen}) for $A$ can be simplified:
\begin{eqnarray}
A &=& \frac{G_{\rm F}}{\sqrt{2}} \, N_B \left[ 1 - 
4 \, \sin^2 \theta_{\rm W} \, Y_e + 2 \, Y_{\nu_e} - 
\frac{8}{3} \, \frac{E <E_e>}{m_Z^2} \, Y_e 
\left( 1 - 4 \, \sin^2 \theta_{\rm W} \right) \right.
\nonumber\\
&-& \left. \frac{32}{3} \, \frac{E <E_{\nu_e}>}{m_W^2} \, Y_{\nu_e} 
\right] . 
\label{eq:A_simp}
\end{eqnarray}

The self-energy operator~(\ref{eq:Sigma}) defines the additional 
electron energy which can be written in the plasma rest frame as:
\begin{equation}
\Delta E  = \frac{1}{2} \left(V - A \, \lambda \, v \right)\,, 
\label{eq:Delta_E}
\end{equation}
where $\lambda = -1$ for the left helicity of the electron, and $\lambda = +1$ 
for the right helicity, $v$ is the electron velocity. 

By this means electrons with left and right helicities acquire different 
additional energies in plasma. 
Similarly to the previous analysis for neutrinos, one can guess that the 
process $e_L \to e_R + \gamma$ is possible for ultra-high energy electrons 
only, due to the relative smallness of the helicity-depending energy shift $A$. 
The energy $E_\lambda$ of the ultra-relativistic electron takes the form:
\begin{equation}
E_{\mp 1} \simeq p + \frac{{\bar m}_e^2}{2 p} \pm \frac{A}{2}\,, 
\label{eq:E_+-}
\end{equation}
where the effective electron mass in plasma is introduced, which is defined by:
\begin{equation}
{\bar m}_e^2 = m_e^2 + C \, m_\gamma^2 \, . 
\label{eq:m_eff}
\end{equation}

Given Eq.~(\ref{eq:E_+-}), the energy and momentum conservation law for the 
process $e_L \to e_R + \gamma$ can be written in the form:
\begin{eqnarray}
p + \frac{{\bar m}_e^2}{2 p} + \frac{A (p)}{2} &=&  
p' + \frac{{\bar m}_e^2}{2 p'} - \frac{A (p')}{2} 
+ k + \frac{m_\gamma^2}{2 k}\,, 
\label{eq:E_cons}\\
{\bf p} &=& {\bf p'} + {\bf k}\,.
\label{eq:p_cons}
\end{eqnarray}
For the first step, let us neglect the non-local weak contribution into $A$ 
in~(\ref{eq:A_simp}). It is justified when the energies are not very high. 
In this case, it is easy to obtain the kinematic condition for the 
process to be opened: 
\begin{equation}
E > E_0 = \frac{m_\gamma^2 + 2 \, {\bar m}_e \, m_\gamma}{2 \, A} \,. 
\label{eq:E_0e}
\end{equation}
Similarly to the neutrino case, the numerical analysis gives for 
the threshold energies:

i) for the solar interior:
\begin{equation}
E_0 \simeq 10^{13} \,{\rm MeV} \,; 
\label{eq:E_0e_sun}
\end{equation}

ii) for the interior of a neutron star:
\begin{equation}
E_0 \simeq 10^{7} \,{\rm MeV} \,.
\label{eq:E_0e_NS}
\end{equation}

Evidently, the inclusion of the non-local weak contribution is necessary. 
As the analysis shows, for the solar interior the process 
$e_L \to e_R + \gamma$ is totally forbidden exactly as the process  
$\nu_{eL} \to \nu_{eR} + \gamma$. As for the neutron star interior, 
there exists a window approximately from 10$^7$ MeV to 10$^{10}$ Mev for 
the process to be opened. 

Hence the process $e_L \to e_R + \gamma$ could be realized in the same 
exotic case when an ultra-high energy electron tries to thread a 
neutron star. 

\section{Solution of the Dirac Equation for a Fermion in Medium}
\label{sec:Solution}

For definiteness, we consider the spin-flip process $f_L \to f_R + \gamma$ 
with the plasmon creation, where $f$ could be both neutrino and electron. 
The amplitude of the process contains the bispinor amplitudes with the definite 
helicities, $u_L$ and $u_R '$, of the initial left-handed 
and the final right-handed fermions. We should remind that helicity states do 
not coinside in general with the chirality states, and tend to them in the 
ultra-relativistic limit only. To define the bispinor amplitudes, one should 
write down the modified Dirac equation for a fermion in plasma. In the 
momentum space, the plasma influence is described by the 
self-energy operator $\Sigma_r$
\begin{equation}
\left({\hat{\cal P}} - m - \Sigma_r \right) u^{(\lambda)} = 0\,, 
\label{eq:Dirac}
\end{equation}
where $m$ is the fermion mass. 
The operator $\Sigma_r$ defined in Eq.~(\ref{eq:Sigma}) provides 
the additional fermion energy $\Delta E$ defined in Eq.~(\ref{eq:Delta_E}). 
It is well known that 
the additional energy in plasma modifies the phase of the de Broglie 
wave of the fermion:
\begin{equation}
\psi (x) \sim {\mathrm{e}}^{- {\mathrm{i}} ({\cal P} x)}\,, \quad 
{\cal P}^\alpha = \left(E + \frac{1}{2} \left(V - A \, \lambda \, v \right), 
{\bf p} \right)\,. 
\label{eq:deBroglie}
\end{equation}

Substituting~(\ref{eq:deBroglie}) into~(\ref{eq:Dirac}), one can rewrite the 
Dirac equation in the plasma rest frame as follows:
\begin{equation}
\left[{\hat{p}} - m - \frac{A}{2} \, \gamma_0 \left(\lambda \, v - 
\gamma_5 \right) \right] u^{(\lambda)} = 0\,. 
\label{eq:Dirac2}
\end{equation}
It is worthwhile to note that the relatively large value $V$ is exactly 
cancelled here and it does not influence the bispinor amplitude 
$u^{(\lambda)}$. 

The solution of Eq.~(\ref{eq:Dirac2}) obtained in the natural approximation 
$A \ll E$, is 
\begin{equation}
u^{(\lambda)} \simeq \left( 1 - \frac{m A}{4 E^2} \, \gamma_0 \gamma_5 \right) 
u^{(\lambda)}_0 \,. 
\label{eq:u^lambda}
\end{equation}
where $u^{(\lambda)}_0$ is the solution with definite helicity of the Dirac 
equation in vacuum, $\left({\hat{p}} - m \right) u^{(\lambda)}_0 = 0$. 

It is seen from Eq.~(\ref{eq:u^lambda}) that in the ultra-relativistic limit 
the deviation of the solution in plasma from the vacuum one contains an 
additional suppressing factor $m/E \ll 1$. Thus, one can use the vacuum 
solution with a great accuracy. 

We have to note that in 
Refs.~\cite{StuTerPLB05,GriStuTerGC05,StuJPA06}, 
\cite{0611100} -- \cite{0611128} 
the authors claimed to 
obtain the exact solution of the modified Dirac equation for a fermion 
in plasma. However, it can be easily seen that at least in one case their 
solution is simply incorrect. Really, for the interior of a neutron star, 
the additional neutrino energy $W \sim 10 $ eV can exceed the momenta 
of soft neutrinos, $p \sim 1$ eV. In this case the bispinor (5) in 
Ref.~\cite{StuJPA06} for the right-handed massless neutrino is not the 
solution of the modified Dirac equation (4) there. Moreover, according to 
Eq. (6) of Ref.~\cite{StuJPA06} such right-handed neutrinos acquire an 
additional energy in plasma, in spite of being sterile. 

\section{Ultra-High Energy Neutrino Threads a Star}

Here, we consider the above-mentioned possibility of the radiative spin-flip 
ultra-high energy neutrino transition 
$\nu_L \to \nu_R + \gamma$ when the neutrino threads a star. 
Obviously it could have only a methodical meaning. 
For these purposes, we present a correct calculation of the process 
width,\cite{Kuznetsov:2006c} with some details.\cite{Kuznetsov:2006d}

A neutrino having a magnetic moment $\mu_\nu$ interacts with photons, and the 
Lagrangian of this interaction is
\begin{equation}
{\cal L} = - \frac{{\mathrm{i}} \, \mu_\nu}{2} \left( \bar \nu \sigma_{\alpha \beta} 
\nu \right) F^{\alpha \beta} \,,
\label{eq:L}
\end{equation}
where $\sigma_{\alpha \beta} = (1/2)\, (\gamma_\alpha \gamma_\beta - 
\gamma_\beta \gamma_\alpha)$, and 
$F^{\alpha \beta}$ is the tensor of the photon electromagnetic field. 

With the Lagrangian~(\ref{eq:L}), one readily obtains the 
amplitude of the process $\nu_L \to \nu_R \gamma^{(\xi)}$ with a 
creation of the plasmon with the four-momentum $q^\alpha = (\omega, {\bf k})$ 
and the polarization $\xi$:
\begin{equation}
{\cal M}^{(\xi)} = \mu_\nu \left({\bar u}_R ' \,\hat q 
\,{\hat \varepsilon}^{(\xi)} 
\, u_L \right), 
\label{eq:M}
\end{equation}
where $u_L$ and $u_R '$ are the bispinor amplitudes for the initial left-handed 
and the final right-handed neutrinos. 

The amplitude~(\ref{eq:M}) squared
\begin{equation}
\left|{\cal M}^{(\xi)} \right|^2 = \mu_\nu^2 \, {\rm Tr} \, \left[\rho_L (p) \, 
{\hat \varepsilon}^{(\xi)} \,\hat q \, \rho_R (p ') \,\hat q 
\,{\hat \varepsilon}^{(\xi)} \right], 
\label{eq:M^2b}
\end{equation}
with the neutrino density matrices substituted, 
\begin{equation}
\rho_L (p) = u_L {\bar u}_L = 
\frac{1}{2} \, \hat p \, (1 + \gamma_5) \,, \quad 
\rho_R (p ') = u_R ' {\bar u}_R ' = 
\frac{1}{2} \, \hat{p '} \, (1 - \gamma_5) \,, 
\label{eq:dens_matr}
\end{equation}
is:
\begin{equation}
\left|{\cal M}^{(\xi)} \right|^2 = 2 \, \mu_\nu^2 \, \left[ 2 (p q)\, (p ' q) 
- q^2 \, (p p ') - 2 \, q^2 \left( p \, \varepsilon^{(\xi)}\right) \left( p ' 
\varepsilon^{(\xi)}\right) 
\right]. 
\label{eq:M^2c}
\end{equation}
Using the energy-momentum conservation law and keeping in mind that 
$E > E_0 \gg W$, one obtains:
\begin{equation}
(p q) = W \,(E - \omega) + \frac{m_\gamma^2}{2}\,, \quad  
(p ' q) = W \,E - \frac{m_\gamma^2}{2} \,, \quad 
(p p ') = W \,\omega - \frac{m_\gamma^2}{2} \,. 
\label{eq:pq}
\end{equation}
Substituting Eqs.~(\ref{eq:pq}) into Eq.~(\ref{eq:M^2c}) and summarizing over 
the transversal plasmon polarizations:
\begin{equation}
\sum\limits_{\xi} \left( p \, \varepsilon^{(\xi)}\right) 
\left( p ' \varepsilon^{(\xi)}\right) = E^2 \, \sin^2 \theta \,,
\qquad
\sum\limits_{\xi} \left|{\cal M}^{(\xi)} \right|^2 = \left|{\cal M} \right|^2\,,
\label{eq:sum}
\end{equation}
where $\theta$ is the angle between the initial neutrino momentum ${\bf p}$ 
and the plasmon momentum ${\bf k}$, one readily obtains the amplitude squared 
in the form:
\begin{equation}
|{\cal M}|^2 = 4 \, \mu_\nu^2 \, E^2 \left[ 2 \, W^2 \left(1 - \frac{\omega}{E} 
\right) - m_\gamma^2 \, \sin^2 \theta \right].
\label{eq:M^2}
\end{equation}
We should note that this expression presented in our 
Ref.~\cite{Kuznetsov:2006c} came under criticism in 
Ref.~\cite{Grigoriev:2006b} where it was declared not to be 
positively-defined. One can see that the plasmon mass $m_\gamma$ 
in the second negative term is much greater than the additional neutrino 
energy $W$. 
However, one should wonder what is the restriction on the $\theta$ angle. 
Really, from the energy-momentum conservation law, the $\theta$ angle can be 
expressed in terms of the energy $\omega$ of the 
plasmon to be considered relativistic with a good accuracy, as 
follows:
\begin{equation}
\sin^2 \theta \simeq \theta^2 \simeq 
\frac{2 W (\omega - E_0)}{\omega^2} \left(1 - \frac{\omega}{E} 
\right),
\label{eq:theta}
\end{equation}
and the positivity of the amplitude squared becomes manifest:
\begin{equation}
|{\cal M}|^2 = \frac{8 \, \mu_\nu^2 \, E^2 \, W^2}{\omega^2} \left(1 - \frac{\omega}{E} 
\right) \left[ (\omega - E_0)^2 + E_0^2 \right].
\label{eq:M^2pos}
\end{equation}

It should be stressed that discussing ultra-high energy neutrinos, and 
consequently the high plasmon energies, one can consider with a good 
accuracy the plasmon mass $m_\gamma$ as a constant depending on the 
plasma properties only, see Eqs.~(\ref{eq:omega_pl_nr})-(\ref{eq:omega_pl_h}). 
This is in contrast to the left-handed neutrino effective mass squared 
$m_L^2$, which is the dynamical parameter, see Eq.~(\ref{eq:mL}).  

The differential width of the process 
$\nu_L \to \nu_R + \gamma$ is defined as:
\begin{equation}
{\mathrm{d}} \Gamma = \frac{|{\cal M}|^2}{8\, E \, (2 \pi)^2} 
 \, \delta (E + W - E' - \omega) \,
\delta^{(3)} ({\bf p} - {\bf p}' - {\bf k}) \, 
\frac{{\mathrm{d}}^3 p'\, {\mathrm{d}}^3 k}{E'\, \omega} \,,
\label{eq:dGamma}
\end{equation}
where the plasmon energy $\omega$ cannot be taken the vacuum one 
($\omega = |{\bf k}|$), as it was done in the $SL \nu$ 
analysis,\cite{LobStuPLB03} -- \cite{StuJPA06} 
but it is defined by the dispersion 
in plasma, $\omega = \sqrt{{\bf k}^2 + m_\gamma^2}$. 

Performing a partial integration in Eq.~(\ref{eq:dGamma}), one obtains for 
the photon (i.e. transversal plasmon) energy spectrum  
\begin{eqnarray}
{\mathrm{d}} \Gamma = \frac{\alpha}{4} 
\left(\frac{\mu_\nu}{\mu_{\rm B}}\right)^2 
\frac{m_\gamma^2 \, W}{m_e^2} 
\, \frac{1 - x}{\varepsilon \, x^2} 
\left[ \,(\varepsilon \, x - 1 )^2 + 1 \right] 
{\mathrm{d}} x  \qquad \left(\frac{1}{\varepsilon} \leqslant x \leqslant 1 \right) \,,
\label{eq:dGamma2}
\end{eqnarray}
where $\mu_{\rm B} = e/(2 \, m_e)$ is the Bohr magneton, and 
the notations are used $x = \omega/E$, and 
$\varepsilon = E/E_0$. Recall that $E_0 = m_\gamma^2/(2 \, W)$ is 
the threshold neutrino energy for the process to be opened. 

Instead of the photon energy spectrum~(\ref{eq:dGamma2}), one can 
obtain also the spatial distribution of final photons. As the 
analysis of Eq.~(\ref{eq:theta}) shows, all the photons are created inside 
the narrow cone with the opening angle $\theta_0$, 
\begin{equation}
\theta < \theta_0 \simeq \frac{\varepsilon - 1}{\varepsilon} \, 
\frac{W}{m_\gamma} \ll 1 \,.
\label{eq:theta0}
\end{equation}
The distribution of final photons over the $\theta$ angle can be presented in 
the form:
\begin{eqnarray}
&& {\mathrm{d}} \Gamma = \frac{\alpha}{4} 
\left(\frac{\mu_\nu}{\mu_{\rm B}}\right)^2 
\frac{m_\gamma^2 \, W}{m_e^2} \, 
\varepsilon \, (\varepsilon - 1) \, \frac{{\mathrm{d}} y}{\sqrt{1 - y}}
\nonumber\\
&&\times \left\{ \frac{1}{\left[ \varepsilon + 1 - (\varepsilon - 1) 
\sqrt{1 - y}\right]^2}
\left[ 1 - \frac{2}{\varepsilon + 1 - (\varepsilon - 1) \sqrt{1 - y}} 
- \frac{(\varepsilon - 1)^2 y}{2 \varepsilon^2} \right] \right.
\nonumber\\
&&+ \left. \frac{1}{\left[ \varepsilon + 1 + (\varepsilon - 1) 
\sqrt{1 - y}\right]^2}
\left[ 1 - \frac{2}{\varepsilon + 1 + (\varepsilon - 1) \sqrt{1 - y}} 
- \frac{(\varepsilon - 1)^2 y}{2 \varepsilon^2} \right] \right\} ,
\label{eq:dGamma3}
\end{eqnarray}
where $y = \theta^2/\theta_0^2, \, 0 \leqslant y \leqslant 1$. 

Performing the final integration in Eq.~(\ref{eq:dGamma2}) over $x$, 
as well as in Eq.~(\ref{eq:dGamma3}) over $y$, one obtains 
the total width of the process
\begin{eqnarray}
\Gamma &=& \frac{\alpha}{8} 
\left(\frac{\mu_\nu}{\mu_{\rm B}}\right)^2 
\frac{m_\gamma^2 \, W}{m_e^2} \; F (\varepsilon) 
\qquad \left(\varepsilon \geqslant 1 \right)\,,
\nonumber\\
F (\varepsilon) &=& \frac{1}{\varepsilon} \left[(\varepsilon - 1) 
(\varepsilon + 7) - 4 (\varepsilon + 1)\, \ln \varepsilon \right] .
\label{eq:Gamma}
\end{eqnarray}

\begin{figure}[htb]
\centerline{\psfig{file=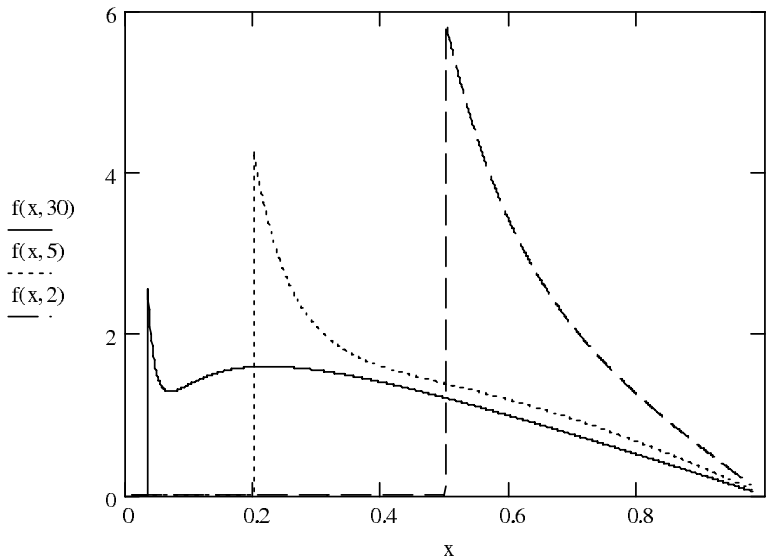,width=100mm}}
\vspace*{8pt}
\caption{The function $f (x, \varepsilon)$
defining the normalized energy spectrum of plasmons from the 
left-handed neutrino decay for different values of the ratio 
$\varepsilon$, 
$\varepsilon = 30$ (solid line), 
$\varepsilon = 5$ (dotted line), 
and $\varepsilon = 2$ (dashed line).
\protect\label{fig:spectrum}}
\end{figure}

\begin{figure}[htb]
\centerline{\psfig{file=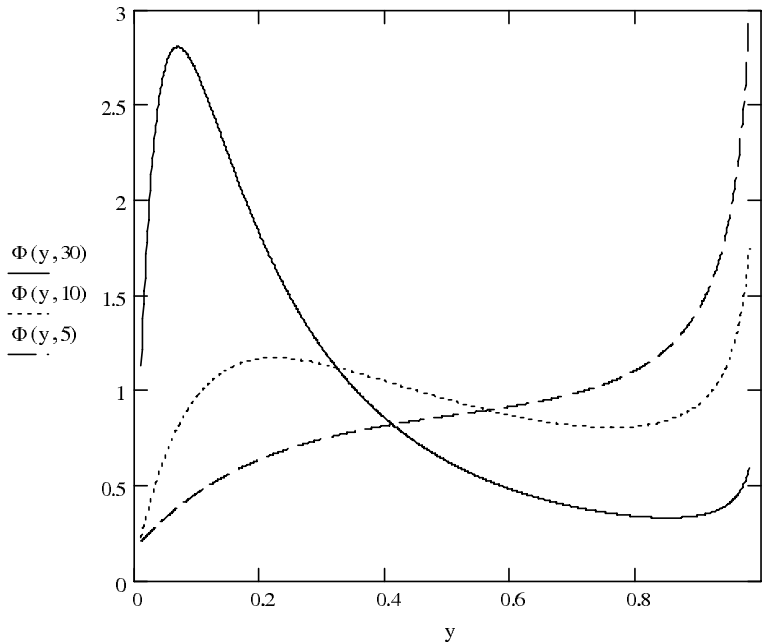,width=100mm}}
\vspace*{8pt}
\caption{The function $\Phi (y, \varepsilon)$
defining the normalized angular distribution of plasmons from the 
left-handed neutrino decay for different values of the ratio 
$\varepsilon$, 
$\varepsilon = 30$ (solid line), 
$\varepsilon = 10$ (dotted line), 
and $\varepsilon = 5$ (dashed line).
\protect\label{fig:angular}}
\end{figure}

In Figs.~\ref{fig:spectrum} and~\ref{fig:angular} the functions 
$f (x, \varepsilon)$ and $\Phi (y, \varepsilon)$ are 
shown, describing the normalized energy spectrum and the normalized 
spatial distribution 
of final plasmons for some values of the ratio $\varepsilon = E/E_0$ of the 
neutrino energy to the threshold neutrino energy:
\begin{equation}
f (x, \varepsilon) = 
\frac{1}{\Gamma} \, \frac{{\mathrm{d}} \Gamma}{{\mathrm{d}} x} \,, 
\quad
\Phi (y, \varepsilon) =
\frac{1}{\Gamma} \, \frac{{\mathrm{d}} \Gamma}{{\mathrm{d}} y} \,, 
\label{eq:f&Phi}
\end{equation}
%

The well-known singularity is seen in Fig.~\ref{fig:angular} 
at $\theta = \theta_0 \, (y = 1)$
which is typical for the angular distribution of a final massive particle 
in the two-particle decay on flight.

It should be noted that in the situation when $W < 0$, and the transition 
$\nu_L \to \nu_R + \gamma$ is forbidden, the crossed channel 
$\nu_R \to \nu_L + \gamma$ becomes kinemalically opened. As the analysis 
shows, the plasmon spectrum and the total decay width are described in 
this case by the same Eqs.~(\ref{eq:dGamma2}) and~(\ref{eq:Gamma}), 
with the only substitution $W \to |W|$. 

To illustrate the extreme weakness of the effect considered, let us 
evaluate numerically the mean free path of an ultra-high energy 
neutrino with respect to the radiative decay, when the neutrino propagates 
through a neutron star. For the typical neutron star parameters, 
$N_B \simeq 10^{38} \, {\rm cm}^{-3}$, $Y_e \simeq 0.05$, 
we obtain
\begin{equation}
L \simeq 10^{19} \, {\rm cm} \times 
\left(\frac{10^{-12}\,\mu_{\rm B}}{\mu_\nu}\right)^2 
\left[F \left(\frac{E}{10 \, {\rm TeV}} 
\right) \right]^{-1} \,, 
\label{eq:path}
\end{equation}
where the neutrino energy $E > E_0$, $E_0 \simeq 10$ TeV is the threshold 
energy for such conditions. 
The mean free path~(\ref{eq:path}) should be compared with the neutron 
star radius $\sim 10^{6}$ cm. 

\section{Ultra-High Energy Electron Spin-Flip Process $e_L \to e_R + \gamma$ 
\\ in Neutron Star}

As was already mentioned, in the papers\cite{0611100} -- \cite{0611128} the 
authors have extended their approach to the so-called 
``spin light of electron'' ($SL e$), $e_L \to e_R + \gamma$. 
However, the same mistake of ignoring the photon dispersion 
in plasma was repeated there. 

According to the analysis performed in Sec.~\ref{sec:Analysis_SL_e}, 
the $SL e$ effect has no place in real astrophysical conditions. 
Here, we consider for methodical purposes the sole possibility when 
an ultra-high energy electron threads a star. 

The process amplitude is caused by the electromagnetic interaction of an 
electron with a photon, and has the form:
\begin{equation}
{\cal M}^{(\xi)} = e \left({\bar e}_R ' \, 
\,{\hat \varepsilon}^{(\xi)} 
\, e_L \right) . 
\label{eq:Meeg}
\end{equation}
After standard but rather cumbersome calculations, one obtains for the 
amplitude squared summarized over the transversal plasmon polarizations $\xi$:
\begin{equation}
\left|{\cal M} \right|^2 
\equiv \sum\limits_{\xi} \left|{\cal M}^{(\xi)} \right|^2  
\simeq 4 \, \pi \, \alpha \, m_e^2 
\left( \frac{E'}{E} + \frac{E}{E'} - 2 \right),
\label{eq:sum_e}
\end{equation}
where $E$ and $E'$ are the enegies of the initial and final electrons. 
Here, $m_e$ is the chiral electron mass which differs from the effective 
mass ${\bar m}_e$, see Eq.~(\ref{eq:m_eff}). 

We note that the amplitude squared of the spin-flip process 
$e_L \to e_R + \gamma$ would be zero if the chiral electron mass $m_e$ is 
equal to zero, because of the chirality conservation in the 
electromagnetic interaction. 

The process probability is calculated by the standard way:
\begin{eqnarray}
\Gamma &=& \frac{\alpha \, m_e^2}{2 \, E^2} \, \int\limits_{E_1}^{E_2} 
{\mathrm{d}} E' \left( \frac{E'}{E} + \frac{E}{E'} - 2 \right),
\label{eq:Gamma_int}\\
\Gamma &\simeq& \frac{\alpha \, m_e^2}{2 \, E} \, 
\left[ \ln \frac{E_2}{E_1} - \frac{E_2 - E_1}{E} 
\left( 2 - \frac{E_2+E_1}{2 E}\right)\right],
\label{eq:Gamma_res}\\
E_{1,2} &=& \frac{E}{2} \left( E + E_0 \, \frac{C}{2 \sqrt{C} + 1}\right)^{-1}
\left[ E + E_0 \, \frac{2 \sqrt{C} - 1}{2 \sqrt{C} + 1} \right.
\nonumber\\
&\mp& \left. 
\sqrt{ \left(E - E_0\right) 
\left( E + E_0 \, \frac{2 \sqrt{C} - 1}{2 \sqrt{C} + 1} \right)} \, \right] ,
\label{eq:E12}\\
E_0 &=& \frac{m_\gamma^2}{2 A} \left( 2 \sqrt{C} + 1 \right).
\label{eq:E_0e2}
\end{eqnarray}
It is seen from Eqs.~(\ref{eq:Gamma_int}) - (\ref{eq:E_0e2}), that $E_0$ is 
the threshold energy. The value $C$ is defined in Eq.~(\ref{eq:V_m_gamma}). 

In the limit $E \gg E_0 \simeq 10^7$ MeV, the expression for the process width 
is simplified essentially:
\begin{equation}
\Gamma \simeq \frac{\alpha \, m_e^2}{2 \, E} \, 
\left( \ln \frac{2 E A}{{\bar m}_e^2} - \frac{3}{2} \right), 
\label{eq:Gamma_app}
\end{equation}
where ${\bar m}_e$ is the electron effective mass, defined in 
Eq.~(\ref{eq:m_eff}). We remind that the probability is not zero only inside 
the window $E_0 < E < E_{\rm{max}}$. The formulas~(\ref{eq:Gamma_int}) 
- (\ref{eq:E_0e2}) were obtained within the approximation 
$E \ll E_{\rm{max}} \simeq 10^{10}$ MeV. 

In contrast to the neutrino spin-flip radiative transition, where the mean 
free path appeared to be extremely large, Eq.~(\ref{eq:Gamma_app}) gives 
the scale of the mean free path for electrons of the order of tens of 
centimeters. However, it is evident that the process considered can not 
compete with the standard electromagnetic processes of the electron 
scattering in dense plasma. 

\section{Conclusion}

In conclusion, we have shown that the effects of the 
``spin light of neutrino'' and ``spin light of electron'' have no place 
in real astrophysical situations for neutrinos and electrons ``living'' in 
plasma, because of the photon dispersion. 
The photon (plasmon) effective mass causes the threshold 
which leaves no room for both processes. For a pure theoretical situation 
when an ultra-high energy neutrino (or electron) threads a star, the total 
probabilities of the processes $\nu_L \to \nu_R + \gamma$ and 
$e_L \to e_R + \gamma$ are calculated with correct taking 
account of the fermion and photon dispersion in plasma.

\section*{Acknowledgments}

We thank S.I. Blinnikov, M.I. Vysotsky, V.A. Novikov, L.B. Okun for useful 
discussion. 

The work was supported in part 
by the Council on Grants by the President of Russian Federation 
for the Support of Young Russian Scientists and Leading Scientific Schools of 
Russian Federation under the Grant No.~NSh-6376.2006.2.


\end{document}